\documentclass[10pt, conference]{IEEEtran}

\usepackage{cite}
\usepackage{amsmath,amssymb,amsfonts}
\usepackage{algorithmic}
\usepackage{graphicx}
\usepackage{textcomp}
\usepackage{mathtools}
\usepackage{ulem}
\usepackage{xcolor}
\usepackage{algorithm,algorithmic}
\usepackage{hyperref}
\usepackage[braket,qm]{qcircuit}
\usepackage{orcidlink}
\usepackage{svg}
\usepackage{comment}
\usepackage{soul}
\usepackage[braket, qm]{qcircuit}

\DeclareMathOperator*{\argmax}{arg\,max}
\DeclareMathOperator*{\MCX}{MCX}
\begin{document}
% \title{Pricing of Rainbow Options With \\Quantum Computers}
\title{Quantum Amplitude Loading for Rainbow \\Options Pricing}

\author{
    \IEEEauthorblockN{
    Francesca Cibrario\orcidlink{0009-0007-8290-4992}\IEEEauthorrefmark{1},
    Or Samimi Golan\IEEEauthorrefmark{2},
    Giacomo Ranieri\orcidlink{0009-0005-0488-2121}\IEEEauthorrefmark{1},
    Emanuele Dri\orcidlink{0000-0002-5144-1514}\IEEEauthorrefmark{3},
    Mattia Ippoliti\orcidlink{0009-0005-9430-2353}\IEEEauthorrefmark{1},
    Ron Cohen\IEEEauthorrefmark{2},\\
    Christian Mattia\IEEEauthorrefmark{1},
    Bartolomeo Montrucchio\IEEEauthorrefmark{3},
    Amir Naveh\IEEEauthorrefmark{2},
    and Davide Corbelletto\orcidlink{0009-0003-8830-2619}\IEEEauthorrefmark{1}
    }
    \\
    \IEEEauthorblockA{
    \IEEEauthorrefmark{1}Intesa Sanpaolo, Torino, Italy\\
    francesca.cibrario@intesasanpaolo.com, giacomo.ranieri@intesasanpaolo.com,\\
    mattia.ippoliti@intesasanpaolo.com, christian.mattia@intesasanpaolo.com, davide.corbelletto@intesasanpaolo.com\\
    \IEEEauthorrefmark{2}Classiq Technologies, Tel Aviv, Israel\\
    orsa@classiq.io, ron@classiq.io, amir@classiq.io\\
    \IEEEauthorrefmark{3}DAUIN, Politecnico di Torino, Torino, Italy\\
    emanuele.dri@polito.it, bartolomeo.montrucchio@polito.it\\
    }
}

\maketitle

% this makes page numbers visible
\thispagestyle{plain}
\pagestyle{plain}

% As a general rule, do not put math, special symbols, or citations in the abstract
\begin{abstract}
This work introduces a novel approach to price rainbow options, a type of path-independent multi-asset derivatives, with quantum computers. Leveraging the Iterative Quantum Amplitude Estimation method, we present an end-to-end quantum circuit implementation, emphasizing efficiency by delaying the transition to price space. Moreover, we analyze two different amplitude loading techniques for handling exponential
functions. Experiments on the IBM QASM simulator validate our quantum pricing model, contributing to the evolving field of quantum finance.
\end{abstract}


%
% For peer-review papers, this IEEEtran command inserts a page break and
\IEEEpeerreviewmaketitle

\section{Introduction}
Quantum computing entails the promise of a paradigm shift in computational technology, offering the potential for solving certain types of problems more efficiently than classical computers. 
One sector poised for significant impact is the financial one, where quantum algorithms hold the potential to benefit tasks like risk analysis \cite{Egger2021, Dri2023}, portfolio optimization \cite{Barkoutsos2020}, and assets pricing\cite{Orus2019, Egger2020, Herman2023}. 

In finance, a crucial aspect of asset pricing pertains to derivatives. 
Derivatives are contracts whose value is contingent upon another source, known as the \textit{underlying}.
The pricing of options, a specific derivative instrument, involves determining the fair market value (discounted payoff) of contracts affording their holders the right, though not the obligation, to buy (call) or sell (put) one or more underlying assets at a predefined strike price by a specified future expiration date (maturity date). 
This process relies on mathematical models, considering variables like current asset prices, time to expiration, volatility, and interest rates. 

Two strategies are commonly used to identify the price of an option: through mathematical models that are solvable analytically or using Monte Carlo simulations.
The former relies on strong assumptions regarding market behavior, while the latter handles more complex scenarios. 
Monte Carlo methods allow the simulation of the evolution of the assets' price given stochastic parameters and provide an estimate of the fair market value exploiting the central limit theorem \cite{boyle1977}. 
The accuracy of the estimate increases with the number of simulations performed, with the confidence interval scaling as $O(1/\sqrt{M})$, where $M$ represents the number of simulations (samples). 
The Monte Carlo simulations approach can be computationally intensive for certain derivatives, such as path-dependent options. 

Quantum computing, therefore, can be a potential advantage when pricing complex options. Using the Amplitude Estimation algorithm, quadratic fewer samples would be required to reach the same result.
Essentially, Amplitude Estimation can estimate a parameter with a convergence rate of $1/M$, where $M$ now is the number of \textit{quantum} samples used. A quantum sample corresponds to an application of the Grover operator, computationally analogous to a classical one.
By lowering the complexity of the Grover operator, the theoretical speedup can in principle be efficiently exploited to achieve reduced execution times.

In this regard, the initial proposals refer to \cite{Rebentrost2018}, in which the authors, for the first time, pioneered a methodology for leveraging the Quantum Amplitude Estimation algorithm in derivative pricing.
The article serves as a starting point for subsequent research efforts that have extended and enhanced the proposed approach.
Specifically, in \cite{Stamatopoulos2020} the authors developed algorithms for various specific classes of options, including vanilla options, multi-asset options, and path-dependent options. Their emphasis lies on the practical implementation of the necessary operators, accompanied by simulation results and an error mitigation scheme designed for execution on real hardware. Notably, their approach stands out for the utilization of the more efficient Amplitude Estimation without Phase Estimation protocol \cite{Suzuki2020}.
More recently, in \cite{chakrabarti2021}, the authors present an upper bound on the resources necessary to achieve a significant quantum advantage in derivative pricing. Moreover, they address some of the challenges of previous approaches by introducing the re-parameterization method to incorporate stochastic processes.

It is important to note that several recent works focused on the intricacies of state preparations \cite{Rattew2022, McArdle, MarinSanchez2023, rattew2023, PhysRevResearch.5.033114}. In order to lower the threshold for achieving a quantum advantage, our work focuses instead on the problem of transitioning to the price space (see Section \ref{sec:background}). This is another module that still presents a high degree of complexity in related works and, to the best of the authors' knowledge, has fewer proposals for optimization in the literature.
Furthermore, to allow an estimate of the complexity of an end-to-end approach the authors decided to focus on a type of options for which, to the best of their knowledge, a suitable quantum algorithm has not yet been proposed: rainbow options.

Rainbow options offer the holder the right to buy or sell an underlying asset at a predetermined price by a specified future date, with the unique feature of multiple underlying assets involved in their valuation \cite{Eales2000} (each underlying can be thought as a color so their sum results in a "rainbow"). 
Rainbow options share a common characteristic with correlation options and basket options in that they all involve multiple underlying securities. 
However, the latter two are tied to a single price determined by those underlying securities. 
In contrast, rainbow options are designed as calls or puts on the best or worst performer among the underlying assets.

In summary, the main contributions of this work are the following:
\begin{itemize}
    \item An end-to-end implementation of a quantum circuit for rainbow options pricing using the Iterative Quantum Amplitude Estimation (IQAE) \cite{Grinko2021} method for estimating the expected payoff, with the advantage of no need for Quantum Phase Estimation \cite{Nielsen2011}. 
    This implementation involved running circuits on a quantum simulator with up to $24$ qubits and two correlated assets, using real-world data.
    \item Additionally, we follow the proposal in \cite{stamatopoulos2023derivative} and adapt it for rainbow options, delaying the transition from the return space to the price space \cite{chakrabarti2021} (see Section \ref{sec:background})  until the payoff calculation phase, raising the need to encode the result of an exponential function inside the amplitude of an auxiliary qubit (exponential amplitude loading). 
This way, we save costly arithmetic computation and combine the transition to price space and amplitude loading in a single efficient block. 
Moreover, we avoid losing precision with digital operations. 
    \item Finally, the paper addresses this crucial aspect of amplitude loading, providing different implementations starting from a generic formulation and narrowing down to the specific case.
    In this regard, two distinct approaches are proposed: \textit{integration} (for strictly positive or strictly negative monotonous functions), and \textit{direct loading} (for exponential functions).
\end{itemize} 
The advantages and nuances of each approach are presented and discussed in depth.

The rest of this paper is organized as follows: Section \ref{sec:background}  introduces the relevant literature and provides the technical foundation for quantum option pricing. Section \ref{sec:methodology} presents our contributions, detailing the methodology behind our algorithm. Section \ref{sec:complexity} contains a detailed analysis of the complexity of the proposed implementation.
Then, Section \ref{sec:experiments} illustrates the experiments conducted and the relative results. Finally, Section \ref{sec:conclusion} contains the final remarks and a summary of our work.

\section{Quantum Option Pricing}
\label{sec:background}
% Qsp (delayed as we did), maybe goes here.

Quantum computing holds significant promise in achieving algorithmic speedup for various computational tasks. 
Specifically, Grover's search offers a theoretical quadratic speedup in the search of unstructured databases \cite{Nielsen2011}.
Beyond database search, Grover's algorithm has been extended to many diverse applications, among which Amplitude Amplification and Estimation. The former enables to reach a theoretical quadratic speedup for estimating expectation values.
This capability holds relevance to problems traditionally addressed using classical Monte Carlo methods \cite{Montanaro2015}.

Option pricing primarily revolves around the computation of an expectation value involving a function applied to one or more stochastic financial underlyings. 
In scenarios extending beyond the Black-Scholes-Merton (BSM) model \cite{Black1973ThePO}, the prevalent approach for such pricing involves resorting to Monte Carlo evaluation techniques.

In recent years, building on this premise and starting from the proposals in \cite{Rebentrost2018}, which introduced the earliest quantum algorithm for valuing European and Asian options, several works dealt with the complexities of quantum option pricing \cite{Stamatopoulos2020, Martin2021, chakrabarti2021}.
In particular, in \cite{Stamatopoulos2020}, the authors describe a structured methodology for pricing several classes of options. 
% The emphasis is on implementing the quantum circuits needed to build the input states and on the operators for Amplitude Estimation.

Summarizing the contributions of these relevant studies, we identify three key components for pricing path-independent options with gate-based quantum computers:
\begin{enumerate}
    %\item represent the probability distribution $\mathbb{P}$ describing the evolution of random variables $\mathbf{X}=$ $\left\{X_1, X_2, \ldots, X_N\right\}$  corresponding to the possible values $S_i$ the underlying assets can take, and the associated probabilities $p_i$ that those values will be realized. This model has to be loaded into a quantum register such that each basis state represents a possible value and its amplitude the corresponding probability. As an example, for an $N-$qubits register, the distribution loading module creates the state: $$
%|\psi\rangle_N=\sum_{i=0}^{2^N-1} \sqrt{p_i}\left|S_i\right\rangle_N

\item Encode the probability distribution of a discrete multivariate random variable $W$ taking values in $\{w_0, .., w_{N-1}\}$ describing the assets' prices at the maturity date. The number of discretized values, denoted as $N$, depends on the precision of the state preparation module and is consequently connected to the number of qubits ($n$) according to the formula $N=2^n$. The resulting encoding for the state preparation is the following:
\begin{equation}\label{eqn:enc_state_preparation}
   \sum_{i=0}^{N-1} \sqrt{p(w_i)}\left|w_i\right\rangle 
\end{equation}
Notably, in related works \cite{chakrabarti2021, Stamatopoulos2020} the assets' price evolution is assumed to be modeled by a Geometric Brownian Motion. Consequently, underlying asset prices adhere to a log-normal distribution.

On this note, in \cite{chakrabarti2021}, two strategies for state preparation are discussed. 
The former involves loading the distribution of asset prices (\textit{price space}), while the latter focuses on loading the distribution of log-returns (\textit{return space}). When asset prices obey a log-normal distribution, then the log-returns are distributed normally.
Moving from the former representation to the latter presents a notable advantage, as it involves preparing the state with a normal distribution instead of a log-normal one. This type of state preparation has been extensively explored in literature, leading to various optimized implementation strategies, as demonstrated in \cite{Rattew2021}. The authors in \cite{chakrabarti2021}, performed the state preparation by loading several standard normal distributions (Gaussians) in parallel and subsequently applying affine transformations to obtain the desired multivariate distribution.
%The noticeable advantage here is that, when asset prices obey a log-normal distribution, then the log-returns are distributed normally. Therefore switching between price space and return space changes from log-normal distribution loading to normal distribution loading, which is usually easier \cite{Rattew2021}. The latter can be loaded by preparing several standard normals (Gaussians) in parallel and then applying affine transformations to get the means and assets' covariance for each Gaussian\cite{chakrabarti2021}. 
% add formulas from points 1 and 2 of alg. 4.2
\item Construct the circuit which computes a scaled version $\tilde{f}$ of the payoff function $f$ inside the amplitude of a target qubit: 
\begin{equation}
\label{eqn:payoff_ae}
\begin{split}  
\sum_{i=0}^{N-1}& \sqrt{(1-\tilde{f}\left(w_i\right)) p\left(w_i\right)}\left|w_i\right\rangle|0\rangle +\\
&\sum_{i=0}^{N-1} \sqrt{\tilde{f}\left(w_i\right)p\left(w_i\right)}\left|w_i\right\rangle|1\rangle
\end{split}
\end{equation}
In \cite{Stamatopoulos2020} the authors encoded directly the payoff inside the amplitude. In contrast, in \cite{chakrabarti2021} the function is encoded into a quantum register before being rotated to the amplitude. It is crucial to emphasize that, when the payoff is defined in terms of prices, it is necessary to adopt a strategy for transitioning to the price space before computing the payoff in case of a return space state preparation.

    \item Calculate the expectation value of the payoff $\mathbb{E}[f]$ using Amplitude Estimation \cite{brassard2002}. The Amplitude Estimation algorithm aims at efficiently estimating $a$ in 
    \begin{equation}
        \mathcal{A}|0\rangle=\sqrt{1-a}\left|\psi_0\right\rangle|0\rangle+\sqrt{a}\left|\psi_1\right\rangle|1\rangle
    \end{equation}
    where $\mathcal{A}$ is a unitary operator while $|\psi_0\rangle$ and $|\psi_1\rangle$ are some normalized states. Thus, $a$ is the probability of measuring $|1\rangle$ in the last qubit. Considering $\mathcal{A}$ as the operator that creates the state in Equation \ref{eqn:payoff_ae}, the value $a$ to estimate is: 

    %For an option with payoff $f$, the $\mathcal{A}$ operator will create the state
    %\begin{equation}
    %\begin{split}  
    %\sum_{\omega_R}^{2^n-1}& \sqrt{(1-\tilde{f}\left(\omega_R\right)) p\left(\omega_R\right)}\left|\omega_R\right\rangle|0\rangle+\\
    %&\sum_{\omega_R}^{2^n-1} \sqrt{\tilde{f}\left(\omega_R\right)p\left(\omega_R\right)}\left|\omega_R\right\rangle|1\rangle
    %\end{split}
    %\end{equation}
    % cambiare S_i
    \begin{equation}
    a=\sum_{i=0}^{N-1} \tilde{f}\left(w_i\right) p\left(w_i\right)=\mathbb{E}[\tilde{f}]
    \end{equation}
    By properly post-processing the estimated $a$ value it is possible to reconstruct the desired expected payoff $\mathbb{E}[f]$ value.
\end{enumerate}
% Threshold for quantum advantage (price/return space), the problem of the exponential, main focus of our work.  
It is worth mentioning that the final step needed for pricing the option is the discounting of the expected value, which is performed classically.

\section{Methodology}
\label{sec:methodology}

\begin{figure*}
    \centering
    \includegraphics[trim={0 17cm 0 4cm}, clip, width=1\linewidth]{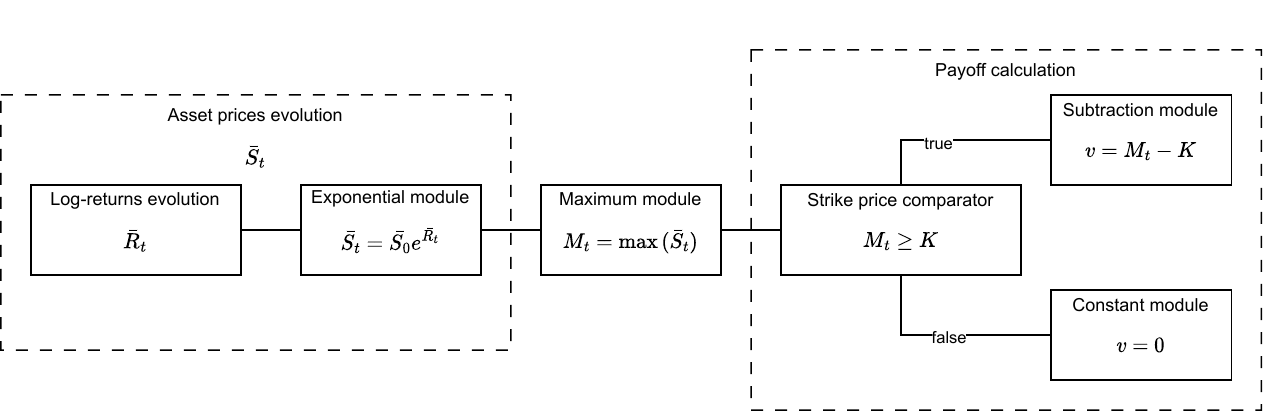}
    \caption{A single evolution of a path in a Monte Carlo Simulation scenario for pricing rainbow options. The approach shows the \textit{price space} version of the pricing algorithm.}
    \label{fig:rainbow_option_graphic}
\end{figure*}

The building blocks of a simulation for pricing call Rainbow Options through a Monte Carlo approach include: 
\textbf{\textit{(i)}} the evolution of assets' prices, \textbf{\textit{(ii)}} selection of the maximum asset price and \textbf{\textit{(iii)}} calculation of the payoff, as illustrated in Figure \ref{fig:rainbow_option_graphic}.
 
\begin{figure*}
    \centering
    \includegraphics[trim={0 9.5cm 0 10cm}, clip, width=1\linewidth]{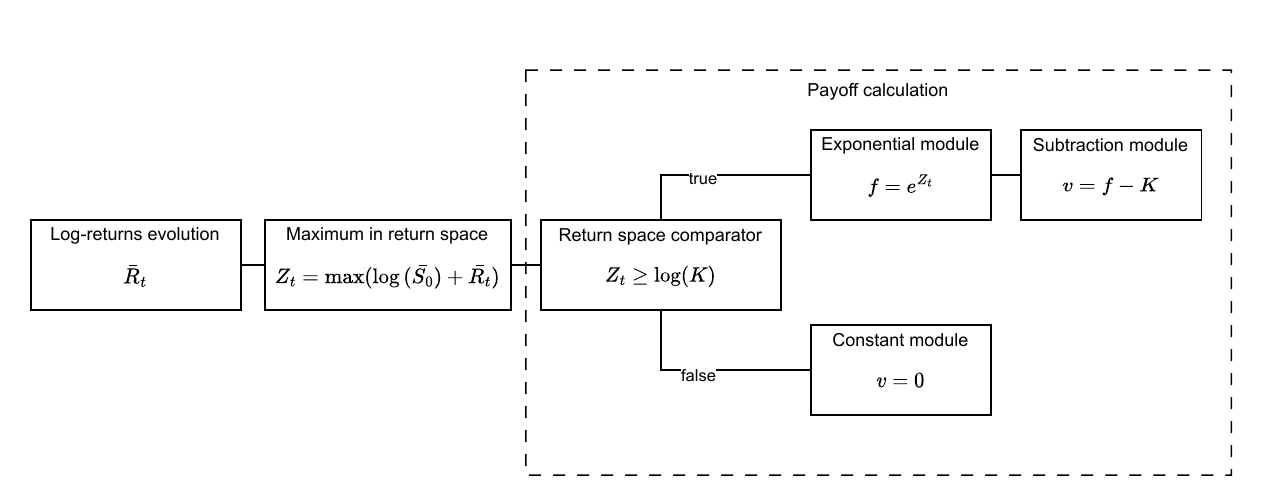}
    \caption{A single evolution of a path in a Monte Carlo Simulation scenario for pricing rainbow options. The approach shows the \textit{return space} version of the pricing algorithm.}
    \label{fig:rainbow_option_quantum_graphic}
\end{figure*}

Under the hypothesis of asset prices being modeled by a Geometric Brownian Motion, at a given $t$ time, the prices follow a multivariate log-normal distribution. Considering $\bar{S_t}$ a vector holding all the asset prices at time $t$, the distributions of log-returns $\bar{R_{t}} = \log{\bar{S_t}/\bar{S_{0}}}$ (where $\bar{S_0}$ denotes the initial prices of the assets), is normally distributed. It is therefore possible to generate standard Gaussian samples and take advantage of the Cholesky decomposition of the correlation matrix, as outlined in \cite{chakrabarti2021}, to establish correlations among the samples.
After log-returns sampling, as depicted in Figure \ref{fig:rainbow_option_graphic}, a transition to the price space becomes necessary to select the asset with the maximum price value ($M_t=\max{(\bar{S_t})}$). This involves executing the operation: $\bar{S_t} = \bar{S_{0}}e^{\bar{R_{t}}}$.
Our strategy consists of delaying this operation until the payoff calculation, as shown in Figure \ref{fig:rainbow_option_quantum_graphic}. 

Therefore, to identify the asset with the maximum value within the return space, the following statement is used:
\begin{equation}
\label{eqn:max}
\argmax(\bar{S_{0}}e^{\bar{R_{t}}}) =
\argmax(\log{(\bar{S_{0}})}+\bar{R_{t}})
\end{equation}
This holds due to the monotonically increasing nature of the exponential function. Therefore the output of the maximum module, in return space, is determined by the quantity:
\begin{equation}
\label{eqn:Zt}
Z_t = \max(\log{(\bar{S_{0}})} + \bar{R_{t}})
\end{equation}

After the maximum module, in the price space, the payoff is calculated as $v(M_t)=\max(M_t-K, 0)$ or equivalently:
\begin{equation}
    v(M_t)=
    \begin{cases}
      M_t - K, & \text{if } M_t \geq K\\
      0, & \text{if } M_t < K
    \end{cases}
\end{equation}
where $K$ is the strike price. 

Since the output of the maximum module in return space is not $M_t$ but the reformulation in Equation \ref{eqn:Zt}, the system in return space becomes:
\begin{equation}
    v(Z_t)=
    \begin{cases}
      e^{Z_t} - K, & \text{if } Z_t \geq \log(K)\\
      0, & \text{if } Z_t < \log(K)
    \end{cases}
\end{equation}
It is therefore possible to move the exponential operation to the payoff calculation phase.

The proposed quantum algorithm implements the flow in Figure \ref{fig:rainbow_option_quantum_graphic}, delaying the exponential module until the payoff calculation phase.
In order to minimize resource usage, a rescaling of all the variables with respect to the first asset is performed. To take into account the rescaling, the payoff calculation can be formalized as follows:
\begin{equation}
\label{eqn:payoff}
    v(z)=
    \begin{cases}
      e^{b \cdot z + b'} - K, & \text{if } z \geq \frac{\log(K) -b'}{b}\\
      0, & \text{if } z < \frac{\log(K) -b'}{b}
    \end{cases}
\end{equation}
with $z = \frac{Z_t-b'}{b}$. The value of the constants $b$ and $b'$, will be addressed in Section \ref{sec:encoding}.
By applying the property of linearity of the mean as it follows:
\begin{equation}\label{eqn:postprocessing}
\begin{split}
&\mathbb{E} \left[ \max{\left(e^{b \cdot z + b'} - K, 0\right)} \right] \\
=\, & \mathbb{E} \left[ \max{\left(e^{b \cdot z}, Ke^{-b'}\right)} \right]e^{b'} - K
\end{split}
\end{equation}
we are able to delay some of the operations needed for the case $z \geq \frac{\log(K) -b'}{b}$, in a post-processing phase, executed classically.
We propose two different strategies for computing $e^{bz}$ using a quantum circuit having $|z\rangle$ encoded inside a state as a fixed point number.

\subsection{Quantum Encoding}
\label{sec:encoding}
The Gaussian state preparation used to load the quantum variables in the return space is encoded as follows:
\begin{equation}
\bigotimes_{i=0}^{N-1}|d_{i}\rangle
\end{equation}
where $N$ is the number of considered assets and $d_i$ represents the state of the $i$-th discrete Gaussian. $|d_i\rangle$ is prepared in the state $\sum_{j=0}^{2^m-1}{\sqrt{p(j)}|j\rangle}$
%\begin{equation}
%\bigotimes_{i=0}^{N-1}{ \left(\sum_{j=0}^{2^m-1}{\sqrt{p(d_{ij})}|d_{ij}\rangle}\right)}
%\end{equation}
where $p(j)$ is the probability of picking the $j$-th sample from a discrete standard Gaussian and $m$ is the number of qubits dedicated to the discretization.
The samples need to be shifted and correlated to obtain the desired multivariate normal distribution. Considering the time delta between the starting date ($t_0$) and the maturity date ($t$), we can express the return value $R_i$ for the $i$-th asset as $R_i = \mu_i + y_i$. Where: 
\begin{itemize}
    \item $\mu_i= (t-t_0)\tilde{\mu}_i$, being $\tilde{\mu}_i$ the expected daily log-return value. It can be estimated by considering the historical time series of log-returns for the $i$-th asset.
    \item  $y_i$ is obtained through the dot product between the matrix $\mathbf{L}$ and the standard multivariate Gaussian sample:
\begin{equation}
    y_i = \Delta x \cdot \sum_kl_{ik}d_k + x_{min} \cdot \sum_k l_{ik}
\end{equation}
$\Delta x$ is the Gaussian discretization step, $x_{min}$ is the lower Gaussian truncation value and $d_k \in [0,2^m-1]$ is the sample taken from the $k$-th standard Gaussian. $l_{ik}$ is the $i,k$ entry of the matrix $\mathbf{L}$, defined as $\mathbf{L}=\mathbf{C}\sqrt{(t-t_0)}$, where $\mathbf{C}$ is the lower triangular matrix obtained by applying the Cholesky decomposition to the historical daily log-returns correlation matrix.
\end{itemize}

We want to normalize the output value of the maximum module ($Z_t$ from Equation \ref{eqn:Zt}) with respect to the first asset. This results in a maximum module that acts on scaled $z_i$ return space values:
$$
z = \max(z_0,.,z_i,.,z_{N-1})
$$
where $z_0 = d_0$ and $z_i$ with $i \neq 0$ are:
\begin{equation}
z_i = \frac{R_i+\log(S^i_0) - \mu_0 - \log(S_0^0) - x_{min}l_{0,0}}{\Delta x l_{0,0}}
\end{equation}
To compute the $z_i$ values and $z$, the maximum of them, an arithmetic module is used to generate the following state:
\begin{equation}
\sum_{D \in [0, 2^m-1]^{N}}{\sqrt{p(D)}}|D\rangle|z(D)\rangle
\end{equation}
%\begin{equation}
%\prod_{i=0}^{N-1}{\sqrt{p(d_i)}}|d_0d_1...d_{N-1}\rangle|z\rangle
%\end{equation}
where $D$ is a N-uple whose $i$-th element, $d_i \in [0,2^m-1]$, represents a sample of the $i$-th standard Gaussian and $p(D) =  \prod_{i=0}^{N-1}{p(d_i)}$. 

Starting from the $z$ value, the inverse operation to apply to obtain $Z_t$ (see Equation \ref{eqn:Zt}) are:
\begin{equation}
Z_t = zl_{0,0}\Delta x + \mu_0 + \log(S^0_0) + x_{min}l_{0,0}
\end{equation}
Recalling Equation \ref{eqn:payoff}, $b$ and $b'$ are therefore: 
$$b = l_{0,0}\Delta x$$
$$b' = \mu_0 + \log(S^0_0) + x_{min} l_{0,0}$$
We can express $z$ as a function of an integer $x$: 
$
z = \frac{x}{2^P}
$
with $P$ the number of fraction places used to encode $z$. We can then rewrite Equation \ref{eqn:payoff}, considering to perform in the post-processing the operations identified in Equation \ref{eqn:postprocessing}, as:
\begin{equation}
f(x) = 
    \begin{cases}
        e^{ax}, & \text{if } \frac{x}{2^P} \geq \frac{\log(K) -b'}{b}\\
      Ke^{-b'}, & \text{if } \frac{x}{2^P} < \frac{\log(K) -b'}{b}
    \end{cases}
\end{equation}
where $a=\frac{b}{2^P}$.

We consider to load a function $\tilde{f}$, that is a modified version of $f$, inside the amplitude using an auxiliary qubit: 
\begin{equation}
\begin{split}
\sum_{i=0}^{2^R-1}{\sqrt{p(x_i)}\sqrt{\tilde{f}(x_i)}|x_i\rangle}|\psi_1\rangle|1\rangle + \\
\sum_{i=0}^{2^R-1}{\sqrt{p(x_i)}\sqrt{1-\tilde{f}(x_i)}|x_i\rangle}|\psi_0\rangle|0\rangle
\end{split}
\end{equation}
where $|\psi_1\rangle$ and $|\psi_0\rangle$ represent additional qubits inside the circuit that have no impact on the Amplitude Estimation outcome, such as the qubits exploited for loading the normal distribution $|D\rangle$. $R$ is the size of $|x\rangle$ register. Amplitude Estimation is used to compute the expected payoff value, collecting the probability of having a $|1\rangle$ in the auxiliary. The post-processing is then needed to obtain  $\mathbb{E}[f]$ from $\mathbb{E}[\tilde{f}]$ and finally $\mathbb{E}[v]$ through the one defined in Equation \ref{eqn:postprocessing}.

To implement $\tilde{f}$, a comparator is used to evaluate the condition $ \frac{x}{2^P} \geq \frac{\log(K) -b'}{b}$ and the outcome is exploited as a control for the two cases modules. 
If the condition is not verified, a rotation is applied to the target qubit to ensure that the amplitude of $|1\rangle$ contains the desired constant value $\tilde{H}$, which is related to $H=Ke^{-b'}$. The needed rotation angle is therefore $\theta=2\arcsin \left({\sqrt{\tilde{H}}}\right)$. 
If otherwise the condition is satisfied, the block performing the exponential amplitude loading (the loading of $\sqrt{\tilde{h}}$, with $\tilde{h}$ a modified version of $h(x)=e^{ax}$, inside the $|1\rangle$ amplitude) is activated.

We faced two approaches for exponential amplitude loading: direct exponential amplitude loading and integration amplitude loading. Both of them require an auxiliary register $|r\rangle$ of the same size ($R$) of $|z\rangle$. 

\subsection{Direct Exponential Amplitude Loading}
\label{sec:direct_loading}
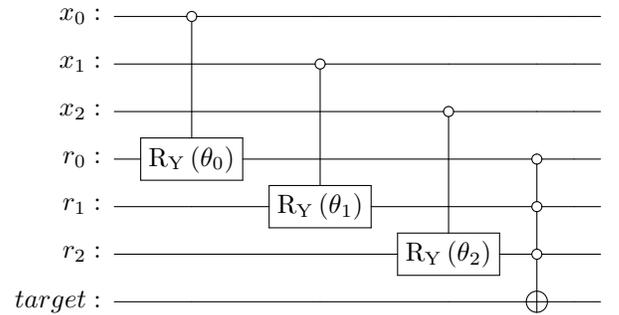
\begin{figure}[h]
\scalebox{1.0}{
\Qcircuit @C=1.0em @R=0.2em @!R { \\
	 	\nghost{{x}_{0} :  } & \lstick{{x}_{0} :  } & \ctrlo{3} & \qw & \qw & \qw & \qw & \qw\\
	 	\nghost{{x}_{1} :  } & \lstick{{x}_{1} :  } & \qw & \ctrlo{3} & \qw & \qw & \qw & \qw\\
	 	\nghost{{x}_{2} :  } & \lstick{{x}_{2} :  } & \qw & \qw & \ctrlo{3} & \qw & \qw & \qw\\
	 	\nghost{{r}_{0} :  } & \lstick{{r}_{0} :  } & \gate{\mathrm{R_Y}\,(\mathrm{\theta_0})} & \qw & \qw & \ctrlo{1} & \qw & \qw\\
	 	\nghost{{r}_{1} :  } & \lstick{{r}_{1} :  } & \qw & \gate{\mathrm{R_Y}\,(\mathrm{\theta_1})} & \qw & \ctrlo{1} & \qw & \qw\\
	 	\nghost{{r}_{2} :  } & \lstick{{r}_{2} :  } & \qw & \qw & \gate{\mathrm{R_Y}\,(\mathrm{\theta_2})} & \ctrlo{1} & \qw & \qw\\
	 	\nghost{{target} :  } & \lstick{{target} :  } & \qw & \qw & \qw & \targ & \qw & \qw\\
\\ }
}
\caption{Circuit implementing the direct exponential amplitude loading. The $|x\rangle$ register controls rotations on the auxiliary $|r\rangle$ register, initialized to $|0\rangle$, to obtain the loading of $e^{-a\hat{x}}$, with $\hat{x}$ binary complement of x, inside the $|0\rangle$ state. An MCX is used to transfer the probability of getting a $|0\rangle$ inside the $|r\rangle$ register to the amplitude of the target qubit, conditioned on the corresponding $|x\rangle$ state probabilities.}
\label{fig:circuit}
\end{figure}

The direct exponential amplitude loading enables encoding the following function in $\tilde{f}$:
\begin{equation}
\tilde{f}(x)=
    \begin{cases}
        e^{-a\hat{x}}, & \text{if } \frac{x}{2^P} \geq \frac{\log(K) -b'}{b}\\
      Ke^{-(b'+ ax_{max})}, & \text{if } \frac{x}{2^P} < \frac{\log(K) -b'}{b}
    \end{cases}
\end{equation}
%$$\tilde{f}(x) = \frac{e^{ax}}{e^{ax_{max}}} = e^{ax-ax_{max}} = {e^{-%a\hat{x}}} 
%$$
where $\hat{x}$ is the binary complement of $x$ ($\hat{x}=x_{max}-x$) and $x_{max}=2^R-1$, the maximum value that can be stored in $|x\rangle$ register. We need to add to the post-processing operation a term:
\begin{equation}
\begin{split}
&\mathbb{E} \left[\max\left(e^{b \cdot z}, Ke^{-b'}\right) \right] e^{b'} - K \\
=\, &\mathbb{E} \left[\max\left(e^{-a\hat{x}}, Ke^{-b'-ax_{max}}\right) \right]e^{b'+ ax_{max}} - K 
\end{split}
\end{equation}
For loading $e^{-a\hat{x}}$, $|r\rangle$ register is initialized to $|0\rangle$ and one controlled rotation for each qubit is performed as shown in Figure \ref{fig:circuit}. The rotations angles are: $\theta_i = 2\arccos \left({\sqrt{e^{-a2^i}}}\right)$. All the probabilities of getting a $|0\rangle$ in the $|r\rangle$ register are then collected by a multi-controlled X  ($\MCX$) gate and stored in the $|1\rangle$ state of a target qubit.

\begin{figure}[ht]
\scalebox{1.0}{
\Qcircuit @C=1.0em @R=0.2em @!R { \\
	 	\nghost{{x}_{0} :  } & \lstick{{x}_{0} :  } & \qw & \multigate{6}{\mathrm{comparator \, (r \leq x)}} & \qw & \qw\\
	 	\nghost{{x}_{1} :  } & \lstick{{x}_{1} :  } & \qw & \ghost{\mathrm{comparator \, (r \leq x)}} & \qw & \qw\\
	 	\nghost{{x}_{2} :  } & \lstick{{x}_{2} :  } & \qw & \ghost{\mathrm{comparator \, (r \leq x)}} & \qw & \qw\\
	 	\nghost{{r}_{0} :  } & \lstick{{r}_{0} :  } & \gate{\mathrm{R_Y}\,(\mathrm{\alpha_0})} & \ghost{\mathrm{comparator \, (r \leq x)}} & \qw & \qw\\
	 	\nghost{{r}_{1} :  } & \lstick{{r}_{1} :  } & \gate{\mathrm{R_Y}\,(\mathrm{\alpha_1})} & \ghost{\mathrm{comparator \, (r \leq x)}} & \qw & \qw\\
	 	\nghost{{r}_{2} :  } & \lstick{{r}_{2} :  } & \gate{\mathrm{R_Y}\,(\mathrm{\alpha_2})} & \ghost{\mathrm{comparator \, (r \leq x)}} & \qw & \qw\\
	 	\nghost{{target} :  } & \lstick{{target} :  } & \qw & \ghost{\mathrm{comparator \, (r \leq x)}} & \qw & \qw\\
\\ }}
\caption{Circuit implementing the integration amplitude loading. The exponential state preparation acts with $R_y$ gates on the $|r\rangle$ register initialized to $|0\rangle$. Then a comparator is used to collect the probability for the $|r\rangle$ register to be lower than the $|x\rangle$ one. In such a way an integration is performed over the state-prepared amplitudes and weighted for the related $|x\rangle$ probabilities.}
\label{fig:integration}
\end{figure}
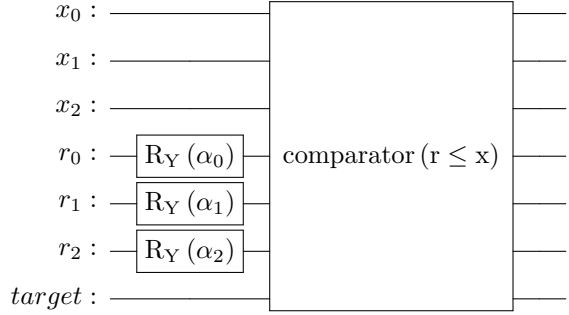
\subsection{Integration Amplitude Loading}
The second method, the integration amplitude loading, is inspired by the work in \cite{krishnakumar2021quantum}, the aim is to use the outcome of a comparator as an integrator to generate $\tilde{h}$ ($\tilde{h}$ is a modified version of $h$). The comparator collects the probabilities $g(r)$ of $|r\rangle$ state until $|r\rangle$ register is lower than $|x\rangle$:
\begin{equation}
\begin{split}
&\sum_{r=0}^{2^R-1}{\sqrt{g(r)}}|x\rangle|r\rangle|r\leq x\rangle \\
=\, &|x\rangle \otimes \left[ \sum_{r=0}^{x}{\sqrt{g(r)}} |r\rangle |1\rangle + \sum_{r=x}^{2^R-1}{\sqrt{g(r)}} |r\rangle |0\rangle \right]
\end{split}
\end{equation}
Collecting the probability to have $r\leq x$ we can define the function:
\begin{equation}
\tilde{h}(x)=\sum_{r=0}^{x}g(r)
\end{equation}
Evaluating the probability to get a $|1\rangle$ results in $\sum_{x = 0}^{2^R-1}{\tilde{h}(x)}$.
To obtain a given function $\tilde{h}$ a proper function $g(r)$ should be chosen.
The $g(r)$ for $r=0$ value must therefore be
$
g(0) = \tilde{h}(0)
$
and for all the other $r$:
$$
g(r) = \tilde{h}(r)-\tilde{h}(r-1)
$$
We first hypothesized to train a variational circuit to perform the state preparation for $g(r)$. 
This approach can be used for every monotonous function $h$ that is strictly positive or strictly negative. 
However, there is a faster method to load the exponential function $h(x)=e^{ax}$ if we consider 
\begin{equation}
\tilde{h}(x) = \frac{h(x+1) - h(0)}{h(x_{max}+1) - h(0)}= \frac{e^{a(x+1)} - 1}{e^{a(x_{max}+1)} - 1}
\end{equation}
getting 
\begin{equation}
g(r)= \frac{e^{ar}}{\sum_{j=0}^{x_{max}} e^{aj}}
\end{equation}
these values can be loaded using a circuit performing an exponential state preparation \cite{Classiq} of unitary depth and no controlled operations. 
The circuit is shown in Figure \ref{fig:integration} where  $\alpha_i=2\arctan{\left(e^{\frac{a2^i}{2}}\right)}$. 
%$\alpha_i = 2\arcsin{\sqrt{\frac{e^{a2^i}}{1+e^{a2^i}}}}$
If we therefore consider:

\begin{equation}
\tilde{f}(x)=
    \begin{cases}
        \frac{e^{a(x+1)} - 1}{e^{a(x_{max}+1)} - 1}, & \text{if } \frac{x}{2^P} \geq \frac{\log(K) -b'}{b}\\
      \frac{Ke^{-b'}}{c} - \frac{e^{-a}}{c}, & \text{if } \frac{x}{2^P} < \frac{\log(K) -b'}{b}
    \end{cases}
\end{equation}
with $c= \frac{e^{a(x_{max}+1)}-1}{e^a}$, the relative updates to the post-processing parameters follow:

\begin{equation}
\begin{split}
\mathbb{E} \left[\max\left(\frac{e^{a(x+1)} - 1}{e^{a(x_{max} +1)}-1}c + \frac{1}{e^a} , Ke^{-b'}\right)\right]  e^{b'} - K \\
=\mathbb{E} \left[\max\left(\frac{e^{a(x+1)} - 1}{e^{a(x_{max} +1)}-1}, \frac{Ke^{-b'}}{c} - \frac{e^{-a}}{c}\right)\right]ce^{b'} \\
+\, e^{b'}e^{-a} - K
\end{split}
\end{equation}

% if needed move
\begin{algorithm}
  \caption{Algorithm}
  \label{alg:algorithm}
  \begin{algorithmic}[1]
    \renewcommand{\algorithmicrequire}{\textbf{Input:}}
    \renewcommand{\algorithmicensure}{\textbf{Output:}}
    \REQUIRE Mean of daily log-return of the $n$ assets,\\
             Desired Strike Price $K$,\\
             Time to expiration $dt$,\\
             Lower triangular matrix from the Cholesky decomposition of log-returns covariance matrix\\
    \FOR {each $n$ assets}
      \STATE Load Gaussian Distribution ($\mu=1,\sigma=0$)
    \ENDFOR
    %\STATE Classically compute the Cholesky decomposition using the log-returns
    \STATE Apply the affine transformation and compute the maximum between all samples
    \STATE Apply the comparator operator between $\frac{(log(K)-b')}{b}$ and the maximum output
    \IF {Integration Method}
        \STATE Load the Exponential State Preparation
        \STATE Apply the comparator between the maximum and the exponential state controlled by the first comparator $|1\rangle$ state and store the comparison result inside the target qubit
    \ENDIF
    \IF {Direct Method}
        \STATE Apply $R_y$ gates controlled by each qubit of the maximum $|0\rangle$ state
        \STATE Apply a $\MCX$ gate controlled by the loaded distribution $|0\rangle$ state and the first comparator $|1\rangle$ state to the target qubit
    \ENDIF
    \STATE Apply the $R_y$ gate controlled by the first comparator $|0\rangle$ state to the target qubit
    % \STATE Uncompute the first comparator
    % \STATE Uncompute the maximum operator
    \STATE Use Quantum Amplitude Estimation to extract the probability of the target qubit to be $|1\rangle$ 
    \ENSURE Expected Payoff for the option by applying the post-processing to the QAE result
  \end{algorithmic} 
\end{algorithm}

\section{Complexity Analysis}
\label{sec:complexity}
Since fault tolerant implementations of the T gate are an expensive resource, minimizing the number of T stages, referred to as the T-depth, is an important target in algorithm development \cite{niemann2019t}.
Our payoff implementation's T-depth is smaller compared to other proposed implementations of the same functionality \cite{chakrabarti2021}.
In this section, we analyze the T-depth of the payoff function for both the direct and integration methods. We do not analyze the assets state preparation module and the arithmetic computing their maximum since they are not the focus of this article. 

Regarding machine precision, we assume the same register size $k$ for each quantum register in the different functions, as they all have the same domain and units.
\begin{comment}
\subsection{Infidelity and Resolution Demand}
\label{subsec:complexity-infi}

The total estimation infidelity ($\tilde{\epsilon}$) in the probability domain, before post-processing, is a summation of the Amplitude Estimation algorithm error $\epsilon_{QAE}$ , of the payoff function infidelity $\epsilon_{payoff}$, and of $\epsilon_{SP}+\epsilon_{max}$ the infidelity of the state preparation and max payoff calculation accordingly.
The scaling factor of the post-processing linearly multiplies this infidelity, which should be taken into account too.
Considering our post-processing strategy, delaying the strike price subtraction, with respect to approaches where it is accounted inside the payoff calculation, the infidelity is affected differently. In particular,  if a precision $\epsilon$ is required on the estimated value, the infidelity of the estimate is affected as:
\begin{equation}
    \tilde{\epsilon} \leq \frac{\epsilon}{S_{max} - K}
\end{equation}
Where $S_{max}$ is the maximum of the discretized values essets and take. 

As for the machine precision, for each quantum register of the different functions, we assume the same register size $k$ as all require the same resolution because they are on the same domain and have the same units.
\end{comment}

The first function that is common to both methods is the $z \geq \frac{\log{(K)} - b'}{b}$ comparator. It compares the $z$ output of the arithmetic maximum module, with a predefined constant $\frac{\log{(K)} - b'}{b}$.  
The Toffoli-depth of a comparator between a quantum register and another quantum register or a constant \cite{draper2004logarithmic} is $2\log_2 (k)+5$. We analyze the Toffoli gate depth, assuming an unlimited number of qubits, and multiply it by a factor of 3 T-depth per Toffoli gate \cite{lee2023t}. 

\begin{equation}
D_{comp}^{T}(k) = D_{comp}^{Toffoli}(k) * 3 = 6\log_2 (k)+15
\end{equation}

For both methods, there is a controlled rotation if the condition is not verified. A controlled $R_y$ gate is composed of two CX gates and two $R_y$ gates. According to Solovay–Kitaev theorem, each $R_y$ gate has a T-depth of:
\begin{equation}
 D_{CR_y}^{T}(\epsilon_{R_y}) =  3 \log_2{(1/\epsilon_{R_y})}  
\end{equation}
where $\epsilon_{R_y}$ is the infidelity of the $R_y$ gate \cite{ross2014optimal}. In the next sections, a detailed analysis of the two methods will be reported. 

\subsection{Direct Exponential Method}
\label{subsec:complexity-exponential}

%The first function of direct exponential mentioned in \ref{subsec:complexity-integration} is the comparator, with the same depth.

For the direct exponential method, after the comparator and the controlled rotation, a direct exponential amplitude loading is implemented with $k$ controlled $R_y$ gates in parallel. %Each controlled $R_y$ gate is composed of two CX gates and two $R_y$ gates.  According to Solovay–Kitaev theorem, each $R_y$ gate has a T-depth $3 log_2(1/\epsilon_{R_y})$ where $\epsilon_{R_y}$ is the infidelity of the $R_y$ gate \cite{ross2014optimal}. 
Demanding payoff infidelity of $\epsilon_{payoff}$ requires stricter $\epsilon_{R_y}$ infidelity for each of the controlled $R_y$ gates.
The payoff infidelity from loading using $k$ controlled $R_y$ gates is $2k \epsilon_{R_y}$, and if considering the the previous $R_y$ gate, the infidelity is $\epsilon_{payoff}=2(k+1)\epsilon_{R_y}$. Therefore, after considering the demand for infidelity, the depth of the exponential loading part and the controlled $R_y$ part is:
%Pi cos()+e - Pi cos() in worst case is k*e
\begin{equation}
\begin{split}
D_{exp}^{T}(\epsilon_{payoff})&+D_{CR_y}^{T}(\epsilon_{payoff})\\
= 2 \cdot D_{CR_y}^{T}(\epsilon_{payoff}) &=
12\log_2{\left(\frac{2(k+1)}{\epsilon_{payoff}}\right)}
\end{split}
\end{equation}

The loading block is followed by a multi-controlled X gate, with $k+1$ control qubits. A reduced T-depth of the MCX is achieved using relative Toffoli gates \cite{maslov2016advantages}. The Toffoli gate depth optimization is obtained through the use of the Classiq platform \cite{mcxClassiq}, resulting in a T-depth of $14\log_3{(n/2)}+5$ where $n$ is the number of controls, equal to $k+1$ in our case.

\begin{equation}
D_{MCX}^{T}(k) = 14\log_3{\left(\frac{k+1}{2}\right)} +5  
\end{equation}

The final T-depth of the payoff function for the exponential method is: 
%6k+15 + 14log_3 \frac{k+1}{2}+5 + 6log_2(1/\epsilon_{exp}) + 6log_2(k/\epsilon_{exp}) 
\begin{equation}
\begin{split}
D_{exponential}^T = D_{comp}^T + D_{exp}^T + D_{MCX}^T + D_{CR_y}^T\\
= 6 \log_2 (k) + 14\log_3{\left(\frac{k+1}{2} \right)}+ 12\log_2{\left(\frac{2(k+1)}{\epsilon_{payoff}}\right)}+20
\end{split}
\end{equation}

\subsection{Integration Method}
\label{subsec:complexity-integration}
\begin{comment}
The first function that is common to both methods is the $z \geq \frac{\log(K) - b'}{b}$ comparator. It compares the $z$ output from the arithmetic maximum function, with a predefined constant $\frac{\log(K) - b'}{b}$.  The Toffoli-depth of a comparator between a quantum register and another quantum register or a constant \cite{draper2004logarithmic} is $2k+5$. We will analyze the Toffoli gate depth, assuming unlimited number of qubits, and multiply it by a factor of 3 T-depth per Toffoli gate \cite{lee2023t}.

$$
D_{comp}^{T}(k) = D_{comp}^{Toffoli}(k) * 3 = 6k+15
$$
\end{comment}
For the integration method, after the comparator and the controlled rotation, we use an integrator, which is a controlled register-to-register comparator. We do not consider the depth of the exponential state preparation before this integrator, since it is parallel to the assets state preparation block and the maximum arithmetic block. This comparator (the integrator) is controlled and therefore implemented by an uncontrolled subtractor, its uncomputation, and a controlled sign check. The sign check, which is a CX gate, becomes a CCX gate in its controlled version. Therefore the T-depth becomes:

\begin{equation}
D_{integ}^{T}(k) = \left(D_{comp}^{Toffoli}(k)+1\right) * 3 =  6 \log_2 (k)+18
\end{equation}

%The last block in the integration method is a controlled $R_y$ gate, which means a T-depth of $6 log_2(2/\epsilon_{payoff})$.

To conclude, when using the integration method, we assume a summation of T-depth from each function:% We do not count the exponential state preparation since it is parallel to the state preparations of the assets and their $max$ calculation. Concluding that, T-depth would be
%6k+15 + 6k+18 + 6log_2(1/\epsilon) 
\begin{equation}
\begin{split}
D_{integration}^T = \;&D_{comp}^T + D_{integ}^T + D_{CR_y}^T\\ 
=\; &12 \log_2 (k)+6\log_2{\left(\frac{2}{\epsilon_{payoff}}\right)}+33
\end{split}
\end{equation}

\subsection{Infidelity and Resolution Demand}
\label{subsec:complexity-infi}

%The total estimation infidelity ($\tilde{\epsilon}$) in the probability domain, before post-processing, is a summation of the Amplitude Estimation algorithm error $\epsilon_{QAE}$, of the payoff function infidelity $\epsilon_{payoff}$, and of $\epsilon_{SP}+\epsilon_{max}$ the infidelity of the state preparation and of the max payoff calculation accordingly.
The total estimation infidelity ($\tilde{\epsilon}$) in the probability domain, before post-processing, is a summation of:
\begin{itemize}
    \item $\epsilon_{QAE}$, the error related to the Amplitude Estimation algorithm
    \item $\epsilon_{payoff}$, the payoff function infidelity
    \item $\epsilon_{GSP}$, the infidelity of the Gaussian state preparation
    \item $\epsilon_{max}$, the infidelity of the arithmetic computing the affine transformation and the maximum
\end{itemize}
The post-processing scaling factor linearly amplifies this infidelity, and it is essential to consider this effect as well. Considering our post-processing strategy, the delay in subtracting the strike price, as opposed to approaches where it is factored into the payoff calculation, results in a different impact on infidelity. 
In particular,  if a precision $\epsilon$ is required on the estimated value, the total infidelity $\tilde{\epsilon}$  should be lower than a quantity that depends on the post-processing. With no delayed subtraction, the infidelity of the estimate is affected as:
\begin{equation}
    \tilde{\epsilon} \leq \frac{\epsilon}{S_{max} - K}
\end{equation}
Where $S_{max}$ is the maximum of the discretized values assets can take. 
The infidelity demand for the integration method is: 
\begin{equation}
    \tilde{\epsilon} \leq \frac{\epsilon}{S_{max} - e^{b'-a}}
\end{equation}
For the direct method, the infidelity is: 
\begin{equation}
    \tilde{\epsilon} \leq \frac{\epsilon}{S_{max}}
\end{equation}
The delayed strike price subtraction has therefore a different impact for the two methods that depends on the input parameters of the pricing problem. 

\section{Experiments}
\label{sec:experiments}
The algorithm was implemented using The Classiq Platform \cite{Classiq}. The platform allowed abstraction by functionally describing the algorithm using high-level building blocks. Each of the parts of the algorithm (state preparation, payoff function, and Amplitude Estimation) was designed through a high-level model. Signed fixed-point quantum variables were used, including the appropriate arithmetic operations over them.

The platform optimizes the quantum program functions by storing multiple implementations for the same functionality and automatically choosing when to uncompute expressions to free auxiliary qubits and re-use them. This compilation approach allowed the scaling of the problem size (number of assets, resolution of distribution and Amplitude Estimation, payoff function complexity) in a given quantum simulation width limit.

\begin{table}[ht]
\centering
\caption{Assets' initial price for the given example}
    \begin{tabular}{l c }
    \hline
    Asset & $S_0$\\
    \hline
    1& \$ 193.97\\
    2& \$ 189.12\\
    \hline
    \end{tabular}
    \label{tab:1}
\end{table}
\begin{table}[ht]
\centering
\caption{Problem parameters for the given example}
    \begin{tabular}{|l |c |c|}
    \hline
    Asset & $\mu$ & Cov\\
    \hline
    1& $5.096 \cdot 10^{-4}$ &$[3.35, 2.57] \cdot 10^{-4}$\\
    2& $6.255 \cdot 10^{-4}$ &$[2.57, 4.18] \cdot 10^{-4}$\\
    \hline 
    \end{tabular}
    \label{tab:2}
\end{table}

The experiments were conducted using the IBM \textit{QASM} noiseless simulator. 
For each experiment, we ran 1000 shots and measured the objective qubit after the IQAE subroutine \cite{Grinko2021}. 
As a proof of concept, we tested our algorithm with two assets whose value at the initial time $t=0$ can be seen in Table \ref{tab:1} and a strike price $K = \$ 190$.  The relative log-returns mean and covariance matrix are visible in Table \ref{tab:2}. Finally, a time to expiration of 250 days (maturity date - starting date) was supposed for the option.

The code used for running these experiments is available in \cite{Cibrario_Quantum_rainbow_option_2024}.

Two qubits were used for each Gaussian distribution. The arithmetic was rounded to one decimal place. An additional qubit was required for the comparator, and the exponential register was sized to match the output of the arithmetic. Finally, one target qubit was used to encode the payoff for the Amplitude Estimation. The Classiq platform also assigned the necessary ancilla qubits, resulting in a total of 24 qubits.

To validate the model, we compared the results obtained using the quantum simulator with the classically computed expected value. 
The classical estimation process considers both the discretization of the Gaussian distributions and the arithmetic precision, in order to achieve a fair comparison.
When using as input the data in Table \ref{tab:1}, the corresponding expected payoff is $\$23.0238$.

\begin{figure}[h]
\centering
\includegraphics[width=0.46\textwidth]{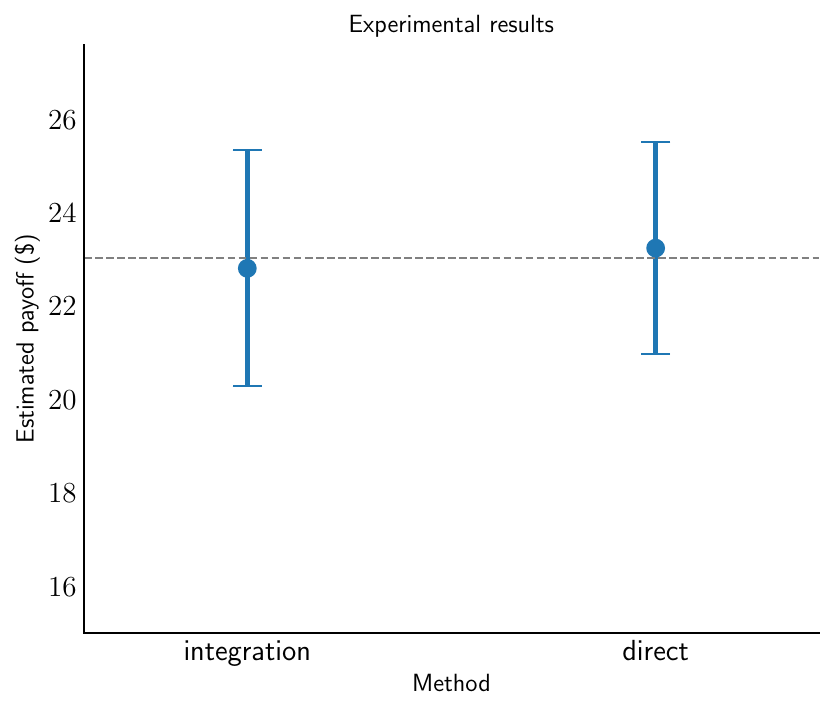}
\caption{Estimated payoff (with relative confidence interval) using the two proposed methodologies for amplitude loading and the IBM QASM simulator. The dotted line represents the classically computed expected value.}
\label{fig:rainbow_plot}
\end{figure}

The simulations were performed using IQAE with a precision $\epsilon = 0.01$ and (for the confidence interval) $\alpha = 0.05$. The results are obtained following the procedure shown in Algorithm \ref{alg:algorithm} and can be seen in Figure \ref{fig:rainbow_plot}. As can be observed, the expected result was in both cases well within the confidence interval of the simulation.

% \section{Results}

\section{Conclusion}
\label{sec:conclusion}
In this paper, we explored the application of quantum computing to the pricing of rainbow options, a derivative product with multiple underlying assets. The traditional methods for pricing options involve computationally intensive Monte Carlo simulations, making them a potential candidate for quantum acceleration through the use of Amplitude Estimation.

Our work builds upon previous research in quantum derivative pricing. We presented an end-to-end implementation for our proposed quantum algorithm, addressing the complexity associated with transitioning to the price space. 
The key contributions of our work include:
\begin{itemize}
    \item Quantum Circuit Implementation: development and implementation of a quantum circuit for rainbow option pricing using the Iterative Quantum Amplitude Estimation (IQAE) method.
    \item Efficient Transition Handling: implementation of a tailor-made strategy (focusing on rainbow options) for delaying the transition from return space to price space until the payoff calculation phase, improving computational efficiency and extending the recent proposal from \cite{stamatopoulos2023derivative}.
    \item Exponential Amplitude Loading: introduction and exploration of two distinct approaches for handling exponential functions in quantum circuits. Both the approaches, namely the integration amplitude loading and the direct amplitude loading, require loading the desired function inside a qubit's amplitude. Direct exponential loading is suitable only for exponential functions, while integration amplitude loading can handle strictly positive or strictly negative monotonous functions. The versatility of the latter extends its potential applications to other algorithms, including HHL \cite{harrow2009}).
    \item Validation Experiments: execution of experiments on a two-asset scenario using the IBM QASM simulator to validate the quantum pricing model by comparing its results with those computed classically.
\end{itemize}

It's important to note that our contributions hold significance for various types of options. As a result, potential future directions could involve addressing more intricate option classes, such as path-dependent options, by leveraging the modules and approaches outlined in this study. Additionally, enhancing the scalability of the existing implementation and conducting small-scale tests on accessible noisy quantum hardware could be considered. Finally, the authors foresee extending the analysis of resource requirements to the entire circuit in the future, aiming to provide more practical assessments of end-to-end implementation needs.


\section*{Acknowledgment}
The authors would like to thank G. Bella for sharing knowledge on the Monte Carlo algorithm for pricing rainbow options.

The authors would also like to thank E. Cornfeld for the fruitful discussion about optimizing the T-depth of multi-controlled X gates and G. Kishony for the prolific discussion about the T-depth counting of various quantum gates.

\bibliographystyle{IEEEtran}
\bibliography{paper}
\end{document}